# CovidCTNet: An Open-Source Deep Learning Approach to Identify Covid-19 Using CT Image

Tahereh Javaheri[†,1], Morteza Homayounfar[†,2], Zohreh Amoozgar[†,3], Reza Reiazi[†,4,5,6], Fatemeh Homayounieh[7], Engy Abbas[8], Azadeh Laali[9], Amir Reza Radmard[10], Mohammad Hadi Gharib[11], Seyed Ali Javad Mousavi[12], Omid Ghaemi[10], Rosa Babaei[13], Hadi Karimi Mobin[13], Mehdi Hosseinzadeh[14,15], Rana Jahanban-Esfahlan[16], Khaled Seidi[16], Mannudeep K. Kalra[7], Guanglan Zhang[1,17], L.T. Chitkushev[1,17], Benjamin Haibe-Kains[4,5,18,19,20], Reza Malekzadeh[21], Reza Rawassizadeh[‡,1,17]

[1] Health Informatics Lab, Metropolitan College, Boston University, Boston, US
[2] Department of Biomedical Engineering, Amirkabir University of Technology, Tehran, Iran
[3] Department of Radiation Oncology, Massachusetts General Hospital, Harvard Medical School, Boston, US
[4] Princess Margaret Cancer Centre, University of Toronto, Toronto, Canada
[5] Department of Medical Biophysics, University of Toronto, Toronto, Canada
[6] Department of Medical Physics, School of Medicine, Iran university of Medical Sciences
[7] Department of Radiology, Massachusetts General Hospital, Harvard Medical School, Boston, US
[8] Joint Department of Medical Imaging, University of Toronto, Toronto, Canada
[9] Department of Infectious Diseases, Firoozgar Hospital, Iran University of Medical Sciences, Tehran, Iran
[10] Department of Radiology, Shariati Hospital, Tehran University of Medical Sciences, Tehran, Iran
[11] Department of Radiology and Gloestan Rheumatology Research Center, Golestan University of Medical Sciences, Gorgan, Iran
[12] Department of Internal Medicine, Iran University of Medical Sciences, Tehran, Iran
[13] Department of Radiology, Iran University of Medical Sciences, Tehran, Iran
[14] Institute of Research and Development, Duy Tan University, Da Nang 550000, Vietnam
[15] Health Management and Economics Research Center, Iran University of Medical Sciences, Tehran, Iran
[16] Department of Medical Biotechnology, School of Advanced Medical Sciences, Tabriz University of Medical Sciences, Tabriz, Iran
[17] Department of Computer Science, Metropolitan College, Boston University, Boston, US
[18] Department of Computer Science, University of Toronto, Toronto, Ontario, Canada
[19] Ontario Institute for Cancer Research, Toronto, Ontario, Canada
[20] Vector Institute for Artificial Intelligence, Toronto, Ontario, Canada
[21] Digestive Disease Research Center, Tehran University of Medical Sciences, Tehran, Iran

[†] Equal contribution in the paper
[‡] Corresponding author, email: rezar@bu.edu

**Abstract**

**Coronavirus disease 2019 (Covid-19) is highly contagious with limited treatment options. Early and accurate diagnosis of Covid-19 is crucial in reducing the spread of the disease and its accompanied mortality. Currently, detection by reverse transcriptase polymerase chain reaction (RT-PCR) is the gold standard of outpatient and inpatient detection of Covid-19. RT-PCR is a rapid method, however, its accuracy in detection is only ~70-75%. Another approved strategy is computed tomography (CT) imaging. CT imaging has a much higher sensitivity of ~80-98%, but similar accuracy of 70%. To enhance the accuracy of CT imaging detection, we developed an open-source set of algorithms called CovidCTNet that successfully differentiates Covid-19 from community-acquired pneumonia (CAP) and other lung diseases. CovidCTNet increases the accuracy of CT imaging detection to 90% compared to radiologists (70%). The model is designed to work with heterogeneous and small sample sizes independent of the CT imaging hardware. In order to facilitate the detection of Covid-19 globally and assist radiologists and physicians in the screening process, we are releasing all algorithms and parametric details in an open-source format. Open-source sharing of our CovidCTNet enables developers to rapidly improve and optimize services, while preserving user privacy and data ownership.**

In the era of communication, the current epidemic of highly contagious Covid-19 (SARS-Cov-2) has negatively impacted the global health, trade and economy. To date the mortality rate of Covid-19 is estimated to be 35-45 times higher than the pandemic influenza, accounting for more than 200,000 deaths [1-4]. Covid-19 has surpassed its predecessors SARS-CoV, and MERS-CoV in morbidity and mortality [5]. Unfortunately, the long-term studies on SARS-CoV, the cause of SARS [6], did not result in finding effective and safe treatments [7]. Lack of effective therapy underlines the importance of early diagnosis, rapid isolation, and strict infection control to minimize the spread of Covid-19.

Currently, diagnosis is mainly based on the patient's medical history, RT-PCR, and CT imaging [8-12]. High error (30-35%) of RT-PCR [8,9,13], lack of distinction between viral contamination versus disease bearing individuals [14] or false negative [15] may have contributed to high prevalence of Covid-19 and the dismal therapeutic outcomes. Here, CT imaging plays a critical role in Covid-19 diagnosis since it not only detects the presence of disease in the lung but also enables identifying the stage of the disease by scoring the CT images [9,16,17]. CT imaging, however, has its own limitations that needs to be addressed. The lack of specificity and the similarities between the lung lesions generated by other types of viral infection or community-acquired pneumonia (CAP) may contribute to misdiagnosis for Covid-19 [18-20]. We hypothesized that the use of robust tools such as machine learning can resolve the CT imaging technical bias and also corrects for human errors [21-26].

An appropriate machine learning strategy for Covid-19 detection should (i) be able to assist radiologists and their staff to rapidly and accurately detect Covid-19, (ii) be compatible with a wide range of image scanning hardware's, and (iii) be user friendly to the medical community without computer-science expertise. In our effort to address the clinical diagnostic needs in the Covid-19 pandemic crisis, we designed a pipeline of deep learning algorithms, CovidCTNet, that is trained on identifying Covid-19 lesions in lung CT images to improve the process of Covid-19 detection.

While deep convolutional neural network approach requires a large dataset [27,28], CovidCTNet by employing BCDU-Net [29] requires only a small sample size for training to achieve accurate detection of Covid-19 by maintaining its robustness to the inevitable biases in the CT image set.

Under institutional review board (IRB) approval, we analyzed the lung CT images to develop a model for accurate detection of Covid-19 from CAP and Control. We acquired CT images from multiple institutions including the publicly available dataset from lung nodule classification (LUNGx) challenge, an archive generated by the University of Chicago [30,31]; and five medical centers in Iran, a country that is highly affiliated with Covid-19. The dataset from Iran was collected from 12 different CT scanner models of five different brands. We collected a heterogeneous dataset to ensure that our algorithm can address the needs of hospitals across the globe, irrespective of the sample size, imaging device (hardware) or the imaging software. Collected CT images covered a variety of image sizes, slice thicknesses, and different configurations of a range of CT scanning devices. Depending on the device and the radiologist decision, the number of scans (e.g., 60, 70, etc.), the image resolution (e.g., 512×512 pixels, 768×768 pixels, etc.), and pixel spaces in the CT images varied. Together these factors allowed us to generate a heterogeneous collection that accounts for differences in CT imaging that exist among the medical community. This broad heterogeneity within the image collection aimed to resolve the potential bias in image analysis towards a specific image quality or types of CT imaging device.

We assessed a dataset consisting of 89,145 slices of all CT scan images from 287 patients. Among this dataset, 109 (32,230 CT slices) patients were infected with Covid-19 with a confirmed RT-PCR, patient's medical history and radiologist diagnosis. The second cohort was 104 (25,699 CT slices) patients infected with CAP or other viral sources with CT images that can be potentially misdiagnosed for Covid-19. Our Control group consists of a cohort of patients 97 (31,216 CT slices) with healthy lungs or other non-Covid-19/non-CAP diseases. Within the Control group 65 cases (20,317 CT slices) were selected from SPIE-AAPM-NCI lung nodule classification challenge dataset [30,32], a heterogeneous dataset that contains lung cancer as well.

In the first step, the CT slices were resampled along three axes (z, y, x) to account for the variety of voxel dimensions among the CT slices (voxel is a single pixel, but in three dimensions). We used the distances of 1×1×1 millimeter for all voxel dimensions. Our method unified CT scans into the same scale and created a resampled dataset from the original dataset, known as resampling to an isomorphic resolution [33] (Fig.1a). In the second preprocessing step, pixel value of the resampled CT images (3D) was optimized to have a proper range of Hounsfield Units (HU). In our dataset, the least dense object such as air takes a value of -1,000. Lung is an organ filled with air and thus acquires a HU value of -700 to -600. Other organs that may interfere with our analysis include water (HU of 0), fat (HU of -90 to -120) soft tissue (HU of 100 to 300), and bone (HU of 300 to 1900). Consequently, we filtered CT slices (2D) to remove non-lung tissue (e.g. skin, bone or scanner bed) that may negatively impact our analysis and to keep only the lung related parts with an HU value ranging from -1000 to 400. Next, a min-max normalization is applied to rescale the -1000 and 400 numerical ranges of pixels to a 0.0 and 1.0 scale (Fig.1b). In the third step of preprocessing, all CT slices of various pixel sizes were resized to a uniform 240×240 pixels on their *x* and *y* dimensions but the number of slices (*z*) remained intact (Fig.1b)

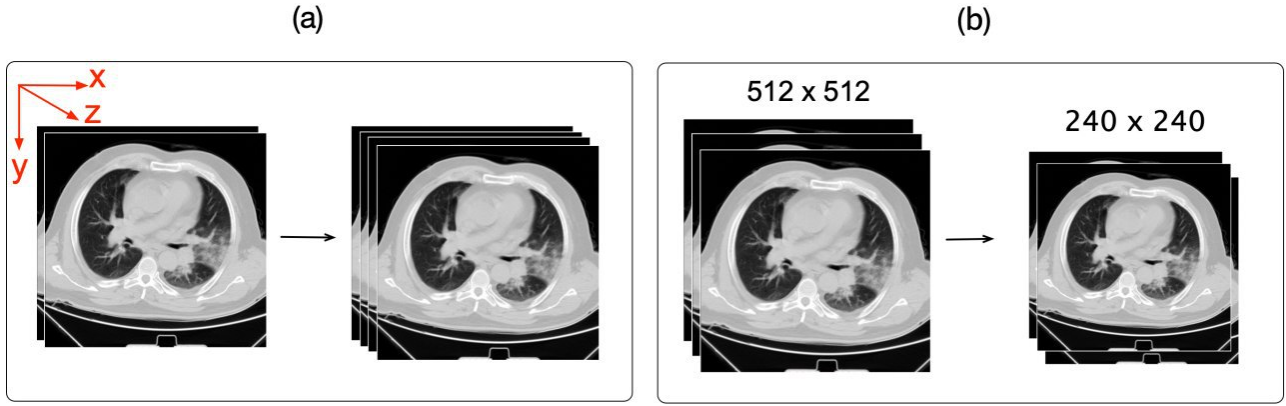

Fig.1| **Schematic representation of the preprocessing phases. a**, Each patient's CT image (3D) was resampled to isomorphic resolution, while x and y are the image coordinates and z represents the number of slices. **b,** All CT slices (2D) with different sizes were resized to have 240×240 pixels on the x and *y* axis, but the z axis that depicts the number of slices remained intact. Here, a 512×512 pixels CT slice is resized into a 240×240 pixels CT slice.

Covid-19 and CAP are associated with lung infection. To model and identify Covid-19 in lung, we generated pseudo-infection anomalies in the CT Control images using the state-of-the-art BCDU-Net [34], which is designed based on the U-Net [35,36], a convolutional network for biomedical image analysis. BCDU-Net has the advantage of having a memory (LSTM cells) allowing the model to remember the structure of the healthy lung. BCDU-Net helped us to increase the accuracy and rate of model convergence by using the initialization of the model that is trained on the Kaggle dataset for lung segmentation [37]. Transfer learning enabled the use of smaller datasets compared to training neural network models from scratch. BCDU-Net focused on lung infection by generating perlin noise [38] (pseudo-infection) and detecting infections. We used the dual function of BCDU-Net in terms of both the anomaly detection and noise cancelation in a unified application. A subset of Control images mixed of noisy and non-noisy were given to BCDU-Net as an input. At the same time, the original Control images of noisy or non-noisy subsets were targeted in the model as output, mimicking the Covid-19 and CAP anomalies in the Control cases (Fig.2a). Feeding the BCDU-Net with CT images without noise, the model learned to identify and remove unnecessary image content, such as heart tissue. In contrast, feeding the BCDU-Net with noisy CT images, the model learned to identify and detect infections or lesions. To this end, we specified the input which is a combination of original healthy CT images and the CT images with noise. We defined the target of our model to be the same input CT images but without noise. Consequently, the output of BCDU-Net is de-noised and reconstructed CT images from the original and noisy CT images. Reconstructing the CT images helped the model to learn the pattern of the Control lung and reconstruct the original Control image as output by noise reduction (de-noised) and infection removal (Fig.2a2-3). Thus, Covid-19 or CAP images could not be reconstructed correctly at this stage. Identifying the Control lung pattern led to recognition of non-Control slices such as Covid-19 or CAP.

In the first step, the training phase (Fig.2a) starts by randomly selecting a dataset of 67 Control patients (22,249 slices) and applying the preprocessing steps on their CT images (Fig.2a1-2). The dataset was divided into two subsets: (i) the original CT images of 11,124 slices (Fig.2a3 right), and (ii) CT images of 11,125 slices with applied perlin noise [38] (Fig.2a3 left). The BCDU-Net model was trained with two noisy and non-noisy subsets of Control images. The trained model was frozen at this step (Fig.2a4).

Next, we preprocessed the entire dataset (excluding the used images in training step) of CT slices including Control, CAP and Covid-19 (Fig.2b5-6) and fed these data into the frozen BCDU-Net model (Fig.2b7). The output of the BCDU-Net is the de-noised CT slices (Fig.2b8). The algorithm subtracted the output of BCDU-Net, the lung slices without infection, (Fig.2b9) from the preprocessed CT slices, infected lung with Covid-19 or CAP (Fig.2b5-6) to acquire the infected areas of lung (Fig.2b9). Because the outcome of subtraction (Fig.2b9) depicted the highlighted infection area (Covid-19 and CAP) without other tissues or artifacts (Fig.2b image in violet colour and 3b), it provided a reliable source for the infection classification as Covid-19 or CAP. In the validation, we observed that the subtraction resulted from original non-infected CT slices versus the output of BCDU-Net was insignificant, confirming the accuracy of detecting noise as an indication for the infected area.

The slices at z axis were concatenated to generate a 3D CT image that was the input of three-dimensional convolutional neural network (CNN) model (Fig.2b10). The outcome of CT slices was resized due to high variation among the number of CT slices for each patient (Fig.2b11). Resizing ensures that all CT images have equal sizes, which is required by CNN to have a unified size ($50\times240\times240$). Here, 50 in the z axis indicates that all the patients' CT slices were resized to 50 slices. These 3D images were already labeled by radiologists as Covid-19, CAP or Control. To implement the classification algorithm, we used CNN. In the final step, the result of Fig.2b11 was fed into the CNN model (Fig.2c12) as a training dataset. In the training phase, the model learned to distinguish Covid-19, CAP and Control. CNN model was then validated by using 24 cases that were selected randomly and were never used before in any of the training and preceding steps. The output of the CNN algorithm is a numerical value that classifies the given patient CT images as Covid-19 or CAP or Control (Fig.2c12). The slices at z axis were concatenated to generate a 3D CT image that was the input of three-dimensional convolutional neural network (CNN) model (Fig.2b10). The outcome of CT slices was resized due to high variation among the number of CT slices for each patient (Fig.2b11). Resizing ensures that all CT images have equal sizes, which is required by CNN to have a unified size ($50\times240\times240$). Here, 50 in the z axis indicates that all the patients' CT slices were resized to 50 slices. These 3D images were already labeled by radiologists as Covid-19, CAP or Control. To implement the classification algorithm, we used CNN. In the final step, the result of Fig.2b11 was fed into the CNN model (Fig.2c12) as a training dataset. In the training phase, the model learned to distinguish Covid-19, CAP and Control. CNN model was then validated by using 24 cases that were selected randomly and were never used before in any of the training and preceding steps. The output of the CNN algorithm is a numerical value that classifies the given patient CT images as Covid-19 or CAP or Control (Fig.2c12).

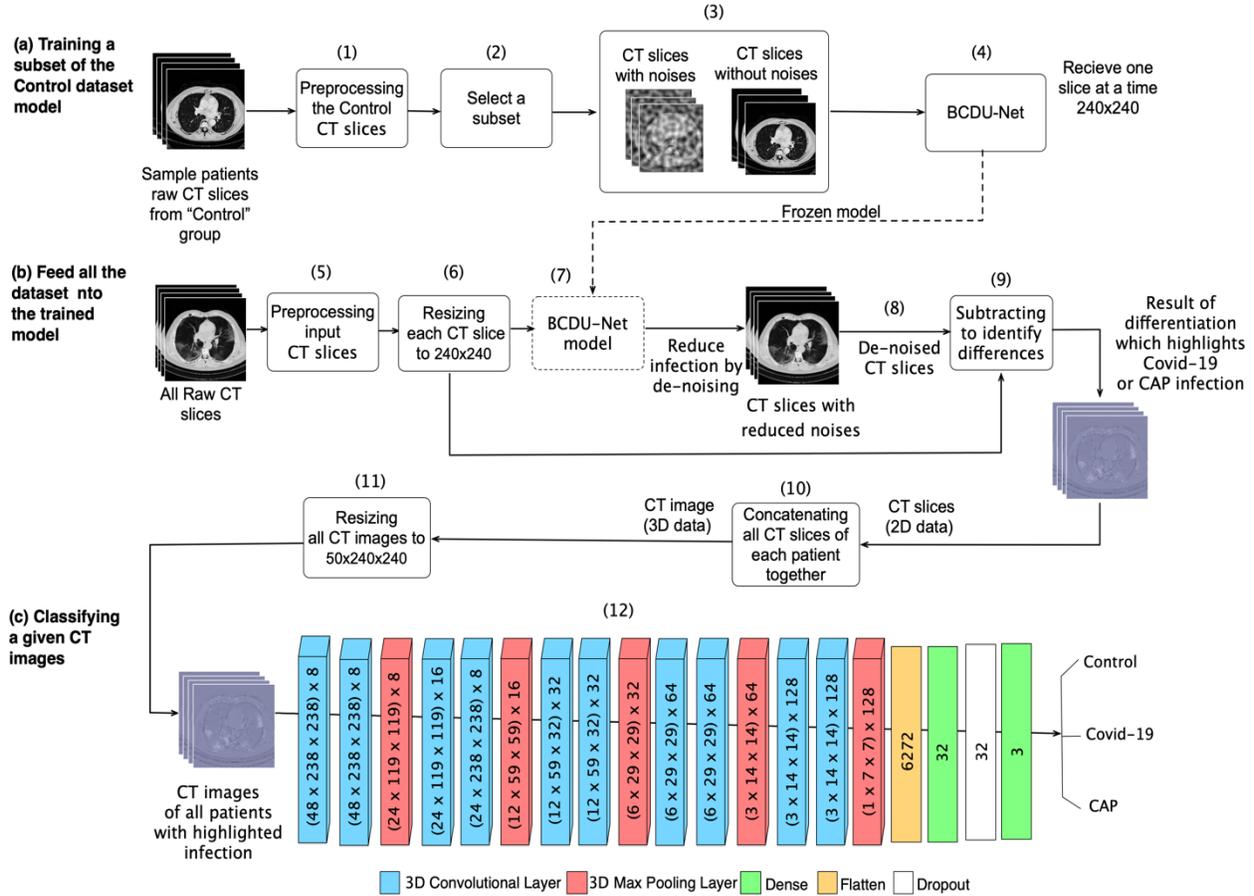

Fig.2| **Multi step pipeline of deep learning algorithm to detect Covid-19 from CT images. a,** Training step of the model for learning the structure of Control CT slices. **b,** Images subtracting and lung reconstructing from CT slices with highlighted Covid-19 or CAP infection (violet colour). The results of step 9 is a 2D image. The slices at z axis concatenated to generate 3D CT image, the input of CNN model. c, CNN model classifies the images that were constructed in the previous stage. To integrate this pipeline into an application the user needs to start from stage (b) and then the CNN algorithm recognizes whether the given CT images of a given patient presents Covid-19, CAP or Control. The number outside the parentheses in CNN model, present the number of channels in the CNN model.

If we feed the original CT images (Fig.3a) or preprocessed CT images into our CNN classification algorithm, the algorithm will not be able to distinguish precisely between Covid-19, CAP and Control lungs in the small dataset. BCDU-Net module (Fig.3) plays a critical role in allowing the detection of infections that has numerous features in a small dataset. Here, the use of a small dataset could compromise model performance while using BCDU-Net, and targeting the infected area at the training phase bypasses the size of dataset. An example of the subtracted data (violet CT slices resulted from Fig.2b11) depicts the infection area in the lung (Fig.3b).

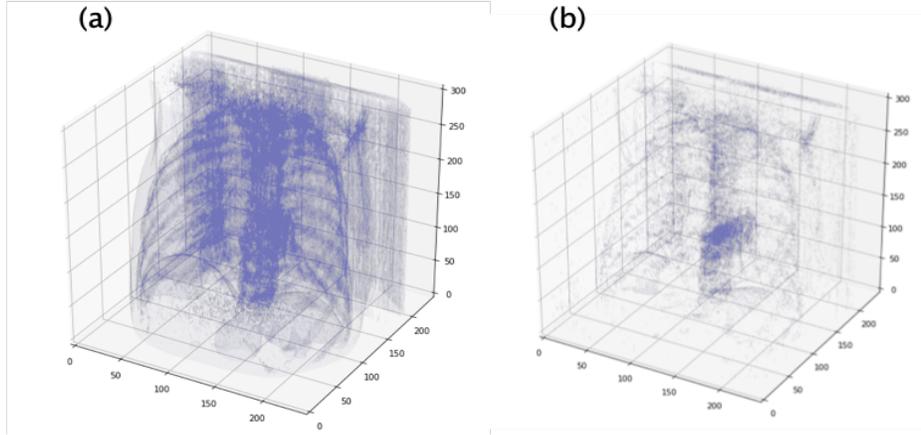

Fig.3| **Schematic representation of BCDU-Net module to detect the infection in CT images. a,** The original CT images visualized in point cloud. **b,** Reconstructed lung image acquired by feeding the CT slices (Fig2b8) into BCDU-Net. The Covid-19 infection area is highlighted in b, which serve as the input of CNN classification model (Fig.2c).

In CNN assessment, the dataset was split in 90% to train the algorithm, and 10% to validate the model in the hold-out. The area under receiver operating characteristics (ROC) curve (AUC) for Covid-19 at the validation phase was 95%, with an accuracy of 91.66% when CNN classified Covid-19 versus non-Covid-19 (two classes). CNN achieved the accuracy of 87.5% when it classified Covid-19 vs CAP and versus Control (three classes). The detection sensitivity of 87.5% and specificity of 94% were recorded for Covid-19 (Fig.4).

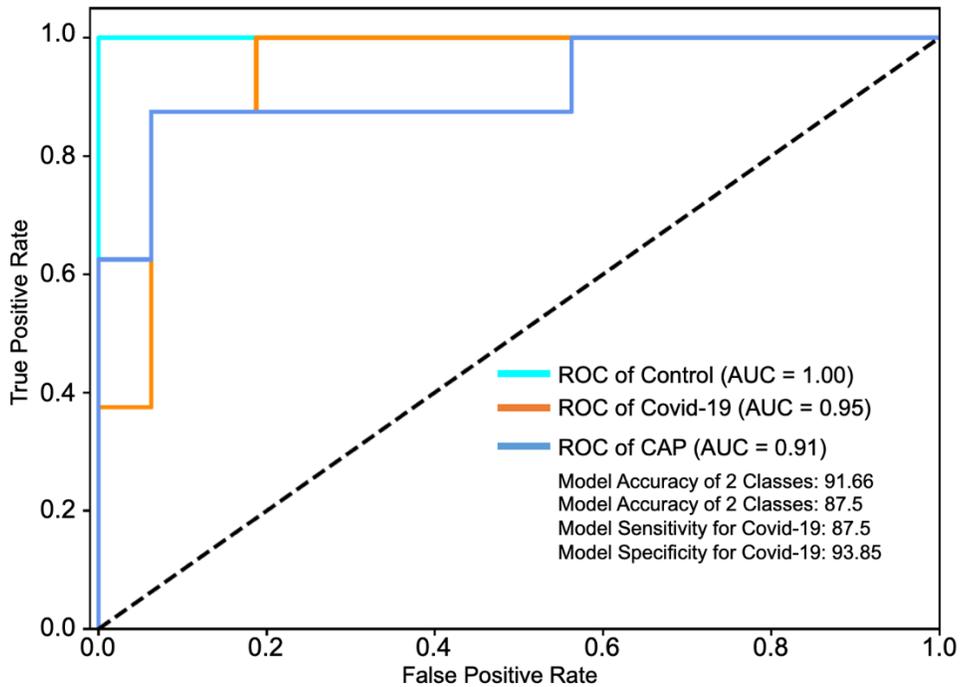

Fig.4| **Performance of CovidCTNet in detecting Control, Covid-19 and CAP.** The model's AUC for Covid-19 detection is 0.95 (n=24 cases). The accuracy, sensitivity and specificity of the model are shown. The model operation in 3 classes demonstrates the detection of all three classes including Covid-19 vs CAP and vs Control and in 2 classes indicates the detection of Covid-19 as one class versus non-Covid-19 (CAP and Control) as second class.

To test the CNN model, an independent dataset consisting of 20 cases mixed of Control, Covid-19 and CAP were assessed using our model and in parallel by four certified and independent radiologists that were not involved in the process of data collection. The average reader performance of four radiologists showed a sensitivity of 79% for Covid-19 and specificity of 82.14%. The CNN, however outperformed the radiologists and achieved Covid-19 detection with sensitivity and specificity of 83% and 92.85% respectively. Radiologists performance accuracy was 81%, while CNN achieved a 90% accuracy when the question was detecting between Covid-19 versus non-Covid-19 (2 classes). When we asked to detect Covid-19 versus CAP versus Control (3 classes), again the machine outperformed the radiologists with an accuracy of 85% compared with human accuracy of 71%. The AUC of the model in Covid-19 detection vs reader test was 93% (Fig.5). The accuracy, sensitivity and specificity of the model showed a significantly higher validity compared to the average of radiologists.

In this study, we developed multiple computational models to accurately detect Covid-19 infection. Due to the high value of patch-based classification [39] we implemented it as one of our models that did not perform as well as CovidCTNet (Extended Data Fig.1). This finding demonstrated the need for decision support tools that assist radiologists in Covid-19 detection.

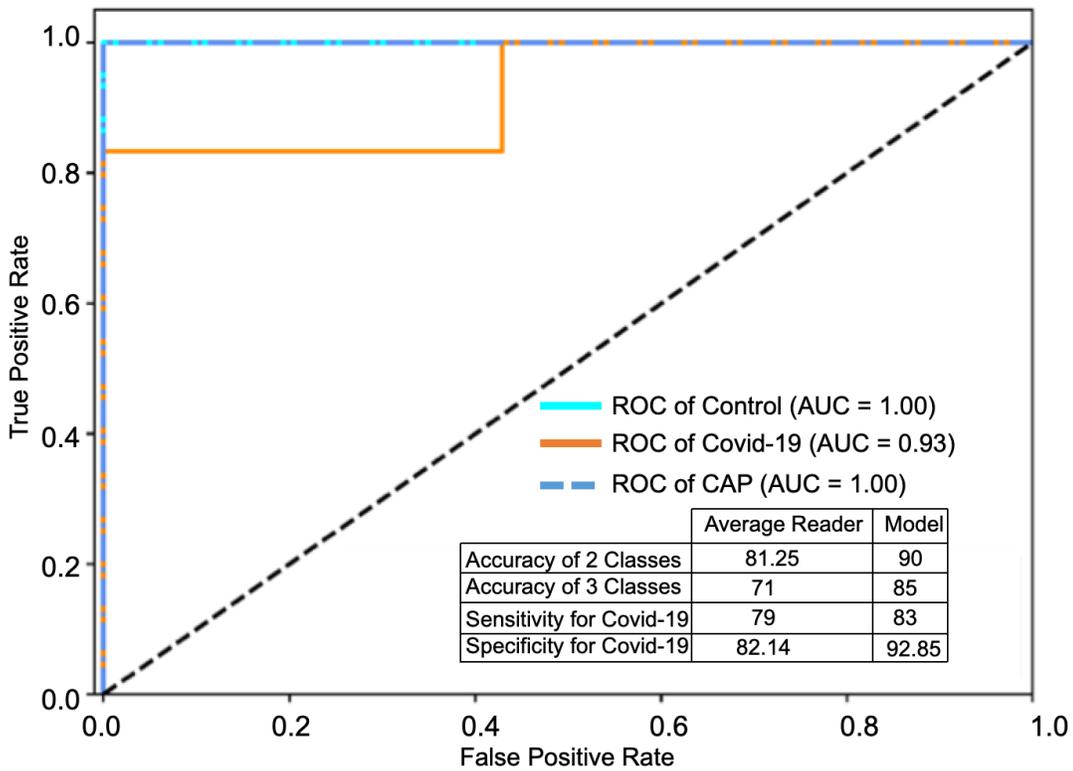

Fig.5| **Comparison of the outcome of CovidCTNet vs reader study.** Performance of model and radiologists (reader) in a pool of chest CT dataset mixed of Control, Covid-19 and CAP. AUC of Covid-19 is 0.93 (n=20 cases). The accuracy, sensitivity and specificity of readers vs model are shown. The model operation in 3 classes demonstrates the detection of all three classes including Covid-19, CAP and Control separately and in 2 classes indicates the detection of Covid-19 as one class versus CAP and Control as second class.

CovidCTNet is an open-source pipeline of a multi-step modelling system (https://gitlab.com/Mohofar/covidctnet.git) that improves the accuracy and consistency of lung screening for Covid-19 detection through a ready-to-use platform with a sensitivity of 83% and accuracy of 90%. In a recent study, it was reported the average sensitivity of radiologists to detect the Covid-19 infection is around 70% [20]. In contrast to recent works [25,40,41,44] our dataset used in this study is smaller and highly heterogeneous. While the broad similarity of patterns and image features of Covid-19 and CAP posed a challenge for algorithm training, the high accuracy of the model indicates the potential for CovidCTNet to be further refined and adapted as a clinical decision support tool. We propose, (i) addition of other samples as the CT scans using in this study are mostly from Iranian patients, (ii) inclusion of the patient's age, gender and medical history to develop a predictive model, (iii) testing the model with a larger number of CT scan databases to further validate and broaden the application of our strategy.

Beyond optimizing and improving Covid-19 detection, CovidCTNet has the potential to significantly impact the clinical workflow and patient care by offering a rapid, inexpensive and accurate methodology to empower healthcare workers during the pandemic and in the future. Importantly, when an infection type is hard to diagnose by the human eye, and when a consensus among radiologists cannot be made, CovidCTNet can be operated as a reliable source of diagnosis. CovidCTNet is the first open-source computational system, designed to assist healthcare professionals in particular radiologists, that allows researchers and developers to adjust and build other applications based on it in a fraction of time [42,43]. Moreover, the addition of patient scans across the course of therapy should enable continuous monitoring of the patient's response to treatment and further development of diagnostic software.

**Online content**
Source data, statements of code and data availability and associated accession codes are available at https://gitlab.com/Mohofar/covidctnet.git. Due to the confidentiality of the codes at the time of submission, the repository is in private mode and can therefore be accessed with request to the corresponding author. Upon the article's acceptance, the code will be available to the public.

**Acknowledgement:** This project was not supported by national or private funding, all authors contributed solely as volunteer. We acknowledge the assistance of Reza Yousefi-Nooraei from University of Rochester, U.S. and Bita Mesgarpour from National Institute for Medical Research Development, Iran for establishing the connections between scientists from different places in the world in a very short time. Besides, we warmly appreciate Ali Babaei Jandaghi from Princess Margaret Cancer Centre, Canada and Nahid Sadighi, from Tehran University of Medical Sciences, Iran for sharing their expertise on CT images annotation. We dedicate our results to all nurses and physicians around the world and especially in Iran, who are risking their lives despite sanctions and international isolation.